\documentclass[twocolumn]{aastex631}

\usepackage{amsmath}
\usepackage{bm}
\usepackage{graphicx}
\usepackage{booktabs}

\newcommand{\msol}{M_{\odot}}

\begin{document}

\title{Constraining first-order phase transition inside neutron stars with application of Bayesian techniques on PSR J0437-4715 NICER data}

\author[0000-0001-6406-1003]{Chun Huang}
\affiliation{Physics Department and McDonnell Center for the Space Sciences  \\
Washington University in St. Louis, 1 Brookings Dr \\
St. Louis; MO, 63130, USA}
\correspondingauthor{Chun Huang}
\email{chun.h@wustl.edu}

\author[0000-0002-0169-4003]{Shashwat Sourav}
\affiliation{Physics Department and McDonnell Center for the Space Sciences  \\
Washington University in St. Louis, 1 Brookings Dr \\
St. Louis; MO, 63130, USA}






\begin{abstract}

Understanding the existence of exotic matter phases and phase transitions within the core of neutron stars is crucial to advancing our knowledge of cold-dense matter physics. Recent multi-messenger observations, including gravitational waves from neutron star mergers and precise X-ray data from NASA’s Neutron Star Interior Composition Explorer (NICER) mission, have significantly constrained the neutron star equation of state (EOS). This study investigates the effects of phase transitions in neutron stars, focusing on NICER’s latest observation of PSR J0437$-$4715. We employ Bayesian inference techniques to evaluate the presence of first-order phase transitions using a piecewise polytropic EOS model. Our analysis incorporates data from multiple NICER sources, to refine constraints on key phase transition parameters, including critical density and transition depth. We find that including data from PSR J0437$-$4715 improves the evidence of phase transitions and tightens the EOS constraints, especially at higher densities. However, Bayes factor analysis only indicates a slight preference for models without phase transitions and current observational precision is insufficient to draw definitive conclusions. In particular, this polytropic model identifies the critical phase transition mass of neutron stars as being close to 1.4 solar masses, which coincides with the approximate mass range of PSR J0437‑4715. This work emphasizes the importance of precise measurements of PSR J0437$-$4715 for deepening our understanding of neutron star interiors and exploring potential new physics at extreme densities.

\end{abstract}

\keywords{stars: neutron -- X-rays: stars -- method: statistical}

\section{Introduction} \label{sec:intro}

Neutron stars are among the most extreme laboratories for cold, dense matter in the Universe, reaching core densities that can exceed those of atomic nuclei by several factors. Under these extreme conditions, unusual phases of matter---such as hyperon or deconfined quarks---may emerge, possibly leading to first-order phase transitions in the stellar interior \citep{2024PhRvD.109f3035C}. Identifying these transitions and understanding how they modify key observables like neutron star masses, radii, and spin evolution is a central challenge in astrophysics. Precise measurements of mass and radius, therefore, are vital for probing the neutron star equation of state (EOS) and discerning whether such exotic forms of matter exist.

Over the past decade, multi-messenger observations have significantly expanded our knowledge of neutron stars. In particular, gravitational-wave detections of neutron star mergers (e.g., GW170817 and GW190425 \citep{LIGOScientific:2017vwq,GW190425})---combined with electromagnetic follow-up---have placed important constraints on the EOS. Complementing these efforts, NASA's Neutron Star Interior Composition Explorer (NICER) mission provides unprecedented precision in measuring both the mass and radius of rapidly rotating millisecond pulsars \citep{Gendreau2016}. Such simultaneous mass--radius determinations offer a unique path to constrain the EOS over a range of densities.

Recent NICER observations have yielded stringent mass--radius measurements for multiple sources. Of particular note, PSR~J0030$+$0451, a neutron star of approximately 1.4\,$M_{\odot}$, placed tight bounds on the EOS and posed new challenges for theoretical models \citep{Riley2019,Miller2019}. Subsequent NICER data for PSR~J0740$+$6620, a $\sim 2\,M_{\odot}$ neutron star, refined our understanding of the EOS at even higher densities \citep{Riley2021,Miller2021,Salmi22}. These measurements are converging on EOS models that appear broadly consistent, although the possibility of phase transitions in the stellar core remains open \citep{Salmi23,Salmi24,Dittmann24}.

Recently, NICER observed PSR~J0437$-$4715 \citep{Choudhury24}, the brightest known millisecond pulsar, which possesses a well-measured mass of $1.44 \pm 0.07\,M_{\odot}$ from radio observation \citep{reardon2024neutronstarmassdistance}. Intriguingly, its radius is markedly different from that of PSR~J0030$+$0451, despite their similar masses. Reconciling this discrepancy may require a first-order phase transition that triggers a steep decrease in radius for only a slight increase in mass. By comparing PSR~J0437$-$4715 to PSR~J0030$+$0451, we can test for such transitions and gain new insights into the EOS at higher densities.

On the theoretical side, Bayesian inference approaches---supplemented by multi-messenger data---have been successful in constraining the likely densities and depth ranges where phase transitions might occur (see eg. \citet{2021PhRvD.103f3026T,PhysRevD.108.094014,PhysRevD.108.043013,Gorda_2023,PhysRevC.107.025801,PhysRevD.88.083013,komoltsev2024firstorderphasetransitionscores}). Nonetheless, definitive observational evidence for first-order phase transitions remains elusive. Motivated by this, we apply Bayesian inference to the new NICER data on PSR~J0437$-$4715 to further assess the plausibility of such transitions. In particular, we employ a polytrope-based meta-model parameterization \citep{2013ApJ...773...11H}, allowing a systematic exploration of the physical conditions that could trigger a first-order phase transition within the neutron star core.

This paper is organized as follows: Section~\ref{bayes} describes our Bayesian framework and prior assumptions. Section~\ref{result} compares models with and without phase transitions, presenting the Bayesian evidence for each scenario. Finally, Section~\ref{discussion} discusses the broader implications of our findings and offers potential directions for future work.

\section{Bayesian Inference Framework}\label{bayes}

In this section, we present the Bayesian framework used to constrain phase transitions in neutron stars. We begin by detailing the chosen prior distributions for our model parameters, based on a polytropic equation of state that includes the possibility of a first-order phase transition. We then describe the data implemented in this analysis, focusing on NICER observations of PSR J0437$-$4715, PSR J0030+0451, and PSR J0740+6620. Our approach involves two comparison scenarios: one analysis incorporates PSR J0437$-$4715, while another excludes it; similarly, we consider models both with and without a phase transition. We evaluate the outcomes using Bayes factors \citep{kass1993bayes,lavine1999bayes,schonbrodt2018bayes} to determine how these observations support models that include phase transitions.

\subsection{Prior Settings}

Bayesian inference has become a highly effective method for determining the internal structure of neutron stars by analyzing their mass and radius measurements \citep{Miller2019, Miller2021, Raaijmakers_2019, Raaijmakers_2020, Raaijmakers_2021}. This approach enables the systematic comparison of various theoretical models and integrates observational data with existing prior information \citep{Hebeler_2013, Greif_2018}. In this study, we utilize a piecewise polytropic (PP) equation of state (EOS) model widely adopted in neutron star research for its ability to capture a diverse range of EOS behaviors, including first-order phase transitions.

To describe the equation of state (EOS), we adopt a three-segment polytropic model incorporating a potential first-order phase transition. This transition is characterized by two adjustable parameters: the transition depth ($\Delta\varepsilon$) and the critical transition density ($\varepsilon_{t}$). These parameters enable us to represent phase changes between the first and second polytropic sections and between the second and third sections. Our approach aligns with the methodologies employed by the NICER collaboration and extends the standard polytropic (PP) model.\citep{Hebeler_2013,Rutherford24}. Normally the standard PP model contains three segment of polytropes. The fact that the three-segment model is primarily used here does not imply that it is sufficient for investigating questions such as quantitative likelihood differences, see more discussions in \citep{PhysRevD.111.034005}. Since the primary goal of our study is to focus on directly comparing with widely-adopted PP model in the literature and investigate the role of phase transition we did not investigate the PP model with more than three segments.

The equation representing our three-piece polytropic EOS is given by:

\begin{equation}
p(\varepsilon)= \begin{cases}
K_{1} \varepsilon^{\gamma_1}, & \varepsilon_c < \varepsilon < \varepsilon_t \\
K_{1} \varepsilon_t^{\gamma_1}, & \varepsilon_t < \varepsilon < \varepsilon_t + \Delta \varepsilon \\
K_{2} (\varepsilon - \varepsilon_t)^{\gamma_2} + K_{1} \varepsilon_t^{\gamma_1}, & \varepsilon_t + \Delta \varepsilon < \varepsilon < \varepsilon_{1} \\
K_{3} (\varepsilon - \varepsilon_{1})^{\gamma_3} + K_{2} (\varepsilon_{1} - \varepsilon_t)^{\gamma_2} \\+ K_{1} \varepsilon_t^{\gamma_1}, & \varepsilon_{1} < \varepsilon < \varepsilon_2
\end{cases}
\end{equation}

Here, $K_1$, $K_2$, and $K_3$ are proportional coefficients for each polytropic segment, and $\gamma_1$, $\gamma_2$, and $\gamma_3$ represent the adiabatic indices. The density at the neutron star's outer crust boundary is fixed at $\varepsilon_{c}=4.3 \times 10^{11} \mathrm{~g} / \mathrm{cm}^3$, while $\varepsilon_{t}$ is the phase transition point, a key parameter explored in this study. The parameter $\Delta\varepsilon$ denotes the transition depth, and $\varepsilon_{1}$ is the transition density between the second and third polytropic segments. The upper density limit, $\varepsilon_2$, is set at 8.3 times the nuclear saturation density, $\varepsilon_{ns}$.

\begin{table}
\centering
\setlength{\tabcolsep}{11mm}
\begin{tabular}{l c}
\hline\hline
\text{EOS parameter} & \text{Prior} \\
\hline
$\gamma_1$ & $\mathcal{U}(1, 4.5)$ \\
$\gamma_2$ & $\mathcal{U}(1, 8)$ \\
$\gamma_3$ & $\mathcal{U}(0.5, 8)$ \\
$\varepsilon_{t}$ & $\mathcal{U}(1.5\varepsilon_{ns}, 8.3\varepsilon_{ns})$ \\
$\Delta\varepsilon$ & $\mathcal{U}(0, 6.8\varepsilon_{ns})$ \\
$\varepsilon_{1}$ & $\mathcal{U}(1.5\varepsilon_{ns}, 8.3\varepsilon_{ns})$ \\
\hline\hline
\end{tabular}
\caption{Summary of prior settings for EOS parameters. $\mathcal{U}$ denotes a uniform (flat) distribution.}
\label{table:1}
\end{table}
The prior distributions for the equation of state (EOS) parameters, presented in Table \ref{table:1}, are informed by earlier research and data from the NICER collaboration, which have investigated similar parameter spaces, see  \cite{Miller2019, Miller2021, Raaijmakers_2019}. These priors are intentionally broad to ensure flexibility within the EOS model while incorporating both theoretical insights and observational constraints. Specifically, uniform priors are assigned to the adiabatic indices $\gamma_1$, $\gamma_2$, and $\gamma_3$, accommodating a wide range of EOS stiffness from soft to very stiff, as highlighted by \cite{Hebeler_2013}. The prior ranges for $\varepsilon_{t}$ and $\Delta\varepsilon$ are established based on nuclear saturation density and the observed characteristics of neutron stars, following the methodology of \cite{Greif_2018}.

In our model, we apply the phase transition constraint exclusively within the density range corresponding to neutron stars with masses between 1.0 and 2.0 $\msol$. This choice is driven by the availability of robust observational data within this mass interval, particularly from pulsars such as PSR J0740+6620. Although phase transitions might occur at higher densities in more massive stars, the scarcity of observational data above 2.0 $\msol$ necessitates focusing our analysis on the most data-rich region. This approach ensures that our Bayesian sampling remains concentrated on the most relevant and informative parameter space, as sampling outside this range would dilute the strength of our constraints by exploring regions where current observations provide little to no information.

By introducing this phase transition model, we aim to fully investigate the constraints that current and new observations, particularly those of PSR J0437$-$4715, can place on the existence and characteristics of phase transitions within neutron stars.

\subsection{Implemented Data}

This work primarily focuses on the potential constraining effects on phase transitions from observation of PSR J0437$-$4715. We restrict our analysis to NICER sources to specifically explore the constraining power of astrophysical observations alone (especially NICER-focused). This approach ensures that all constraints derived from X-ray observation can be independently cross-validated against different sources. The mass and radius measurements inferred from NICER observations of PSR J0030+0451 and PSR J0740+6620 will serve as our comparison group. These measurements are independent, so the likelihood can be written as shown below:

\begin{equation}
\begin{aligned}
& p(\boldsymbol{\theta}, \varepsilon \mid \boldsymbol{d}, \mathcal{M}) \propto p(\boldsymbol{\theta} \mid \mathcal{M}) p(\varepsilon \mid \boldsymbol{\theta}, \mathcal{M}) \\
& \times \prod_j p\left(\mathcal{M}_j, R_j \mid d_{\mathrm{NICER}, \mathrm{j}}\right).
\end{aligned}
\end{equation}

Here, $\boldsymbol{\theta}$ represents the set of parameters describing the neutron star equation of state (EOS), which includes adiabatic indices $\gamma_1$, $\gamma_2$, and $\gamma_3$, as well the transition density $\varepsilon_t$ and transition depth $\Delta \varepsilon$. The symbol $\boldsymbol{d}$ denotes the data set, which consists of the mass and radius measurements of neutron stars, that are inferred from NICER observations. The term $\mathcal{M}$ refers to the model used to describe the neutron star EOS, which incorporates both polytropic segments and the possibility of phase transitions. The likelihood function $p(\boldsymbol{\theta}, \varepsilon | \boldsymbol{d}, \mathcal{M})$ expresses the probability of observing the data $\boldsymbol{d}$ given the parameters $\boldsymbol{\theta}$ and model $\mathcal{M}$, with $\varepsilon$ representing the energy density or other relevant quantities depending on the model. The prior distribution $p(\boldsymbol{\theta} | \mathcal{M})$ reflects the initial beliefs about the parameters before considering the data. The term $p(\varepsilon | \boldsymbol{\theta}, \mathcal{M})$ is the likelihood of the energy density $\varepsilon$ given the parameters $\boldsymbol{\theta}$ and model $\mathcal{M}$. Finally, the product $\prod_j p(\mathcal{M}_j, R_j | d_{\text{NICER}, j})$ represents the likelihood of each NICER measurement for different pulsars, where $\mathcal{M}_j$ and $R_j$ are the mass and radius of the $j$-th neutron star, and $d_{\text{NICER}, j}$ is the corresponding observed data. We will conduct two different inferences: the first using only PSR J0030+0451 and PSR J0740+6620, and the second including the new observation of PSR J0437$-$4715. These results will be compared to inferences made with a model that does not include phase transitions. For the without phase transition (NPT) model, each dataset (with and without PSR J0437$-$4715) will also be analyzed. Comparisons between the same dataset under different models will be made using Bayes factors to determine whether the new observation provides any preference for phase transitions.

For PSR J0740+6620, we use the measurements provided by \cite{Salmi24}, which reported a neutron star mass of approximately 2.0 $\msol$.The phase transition constraining effect from PSR J0030+0451 is discussed based on posteriors from the ST+PDT model for PSR J0030+0451 as presented by \cite{Vinciguerra24}; this model has been consistent with the inferred magnetic field geometry from gamma-ray emission  \citep{2021ApJ...907...63K}. The ST+PDT model offers a plausible and precise estimate for the mass-radius of this source, making it a valuable input for our study. In \cite{Rutherford24}, people also highlight the accuracy of this method for extracting mass-radius posteriors under pulsar timing observations. Different choices of the M-R posterior will naturally affect the results. However, our focus is on the ST+PDT posterior since this model is considered more plausible and consistent with the magnetic field geometry inferred for the gamma-ray emission of this source \cite{2021ApJ...907...63K}, as discussed in \cite{Vinciguerra24}. This geometry is also elaborated on in \cite{chun24hotspot} for detailed modeling of the T-map of this pulsar.

All inferences in this work utilize the \texttt{CompactObject} package developed by the author \citep{EoS_inference,Huang:2024rfg} \footnote{{Zenodo repository of CompactObject package:{\url{https://zenodo.org/records/14181695}}}} . This package is the first open-source, well-documented, full-scope package designed for implementing Bayesian constraints on the neutron star EOS. It accommodates multiple EOS models, including relativistic mean field (RMF) and polytropes. This polytrope model has also been implemented in the \texttt{NEoST} software \cite{Raaijmakers2025}, which focuses on Bayesian inference of the EOS using an agnostic equation-of-state model. Here, the construction of the EOS without a phase transition is adopted from \texttt{NEoST} to keep consistency of model construction. Other studies based on this \emph{CompactObject} package include \cite{Huang:2023grj,Huang:2024rvj}. The nested sampling of this package using the UltraNest\footnote{\url{https://johannesbuchner.github.io/UltraNest/}} package \citep{buchner2021ultranestrobustgeneral}. Fifty thousand live points were utilized for every inference as a base to compare the Bayes evidence; the slice sampler in UltraNest was used, which is well-suited and efficient for high-dimensional sampling. It also ensures consistency in the convergence speed of the sampling process.

\begin{table}[htbp]
\centering
\begin{tabular}{ccc}
\hline\hline
Fitted Model, Without J0437 & ln(Z) & Bayes' Factor\\ 
\hline 
Polytrope, With PT (PT)     & -9.14      & ...\\
Polytrope, Without PT (NPT) & -8.8      & NPT/PT = 1.40\\
\hline 
\hline\hline
Fitted Model, With J0437  & ln(Z) & Bayes' Factor\\ 
\hline 
Polytrope, With PT (PT)     & -11.41      & ...\\
Polytrope, Without PT (NPT) & -10.56     & NPT/PT = 2.34\\
\hline 
\end{tabular}

\caption{This table gives the global log evidence ($\ln Z$), as returned by Ultranest, for the prior models with and without phase transition under two different data sets, with and without PSR J0437$-$4715. PT/NPT denote the Bayes' factor of the PT vs. NPT model. NPT/PT denotes the NPT model compared to the PT one. The choice of Bayes' factor calculation depends on which has more Bayesian evidence.}
\label{Bayes_factor}
\end{table}

\begin{figure*}
    \centering
    \begin{minipage}{0.32\linewidth}
        \centering
        \includegraphics[width=1\linewidth]{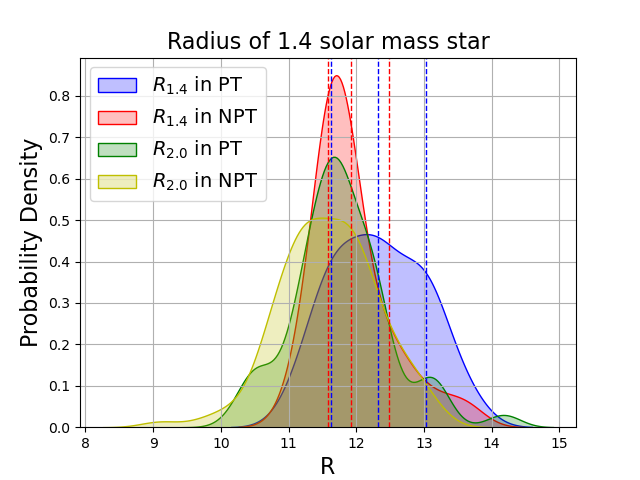}
        \label{critical}
    \end{minipage}%
    \begin{minipage}{0.32\linewidth}
        \centering
        \includegraphics[width=1\linewidth]{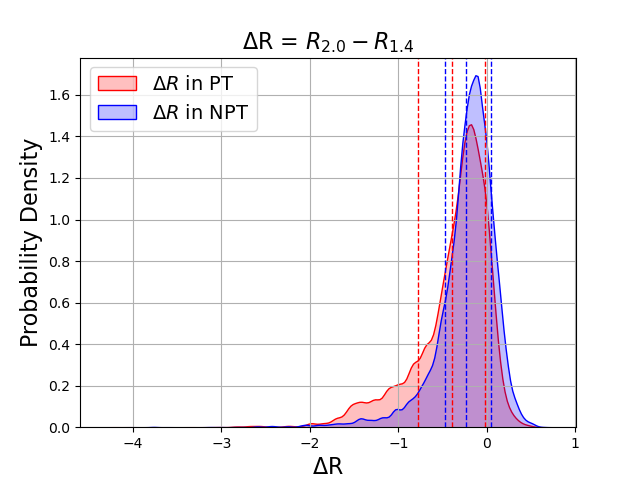}
        \label{Rdiff}
    \end{minipage}%
    \begin{minipage}{0.32\linewidth}
        \centering
        \includegraphics[width=1\linewidth]{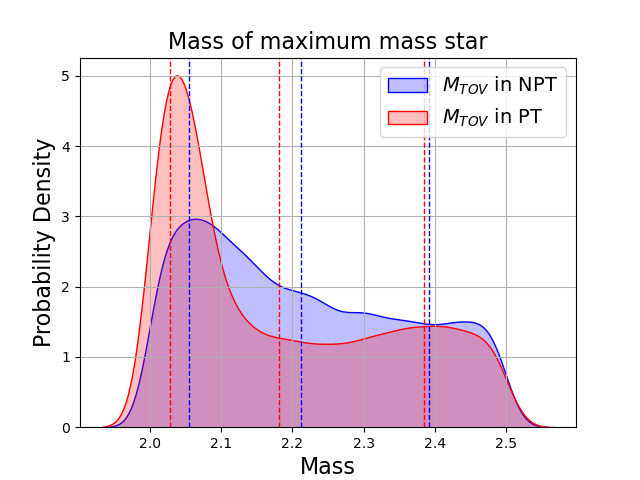}
        \label{mtov}
    \end{minipage}
    \caption{The 1-D distribution of (a) 1.4 $M_{\odot}$ star radius, (b) the radius difference: $\Delta R = R_{2.0} - R_{1.4}$, and (c) Maximum mass ($M_{TOV}$) neutron star in NPT and PT models. The contour levels, from left to right, represent the 16\%, 50\%, and 84\% quantiles of the 1.4 $M_{\odot}$ star radius in NPT and PT models.}
    \label{ig:shiftcombined}
\end{figure*}

\begin{figure*}[htbp]
    \centering
    \includegraphics[scale=0.4]{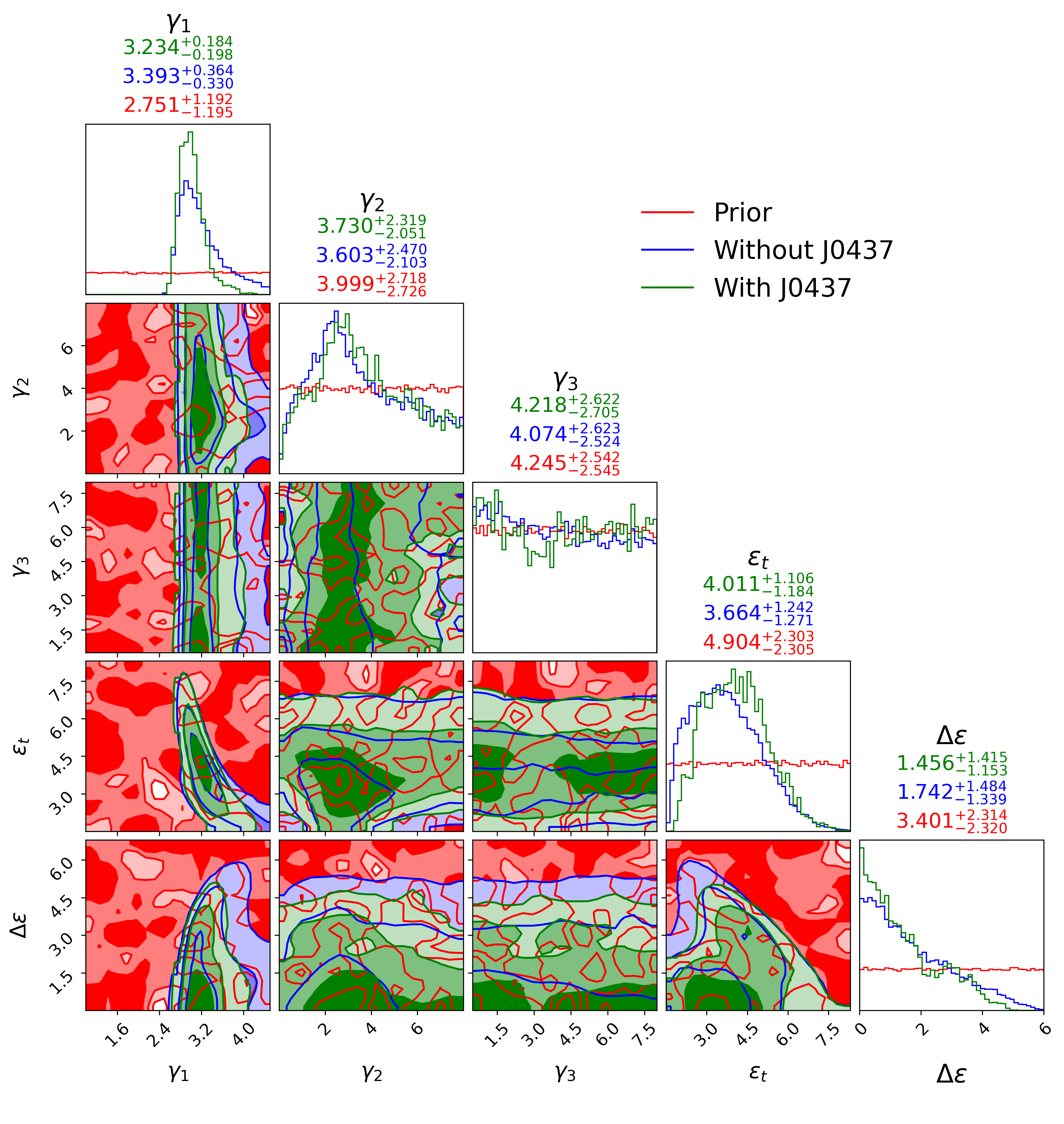}
    \caption{The posterior of the two phase transition parameters and three polytrope indeices after applying constraints to all existing observations with and without including PSR J0437$-$4715. Red is the prior, Blue is the posterior including all the other NICER sources constraints without PSR J0437$-$4715, and green is the posterior with all the  X-ray observation constraint result including PSR J0437$-$4715 . The contour levels in the corner plot, going from deep to light colours, correspond to the 68\%, 84\%, and 98.9\% levels. The dashed line in the 1D corner plots represents the 68\% credible interval, and the title of this plot indicates the median of the distribution as well as the range of the 68\% credible interval. Here $\varepsilon_{t}$ and $\Delta \varepsilon$ are given in saturation density units.}
    \label{prior_post_compare}
\end{figure*}

\begin{figure}[htbp]
    \centering
    \includegraphics[scale=0.4]{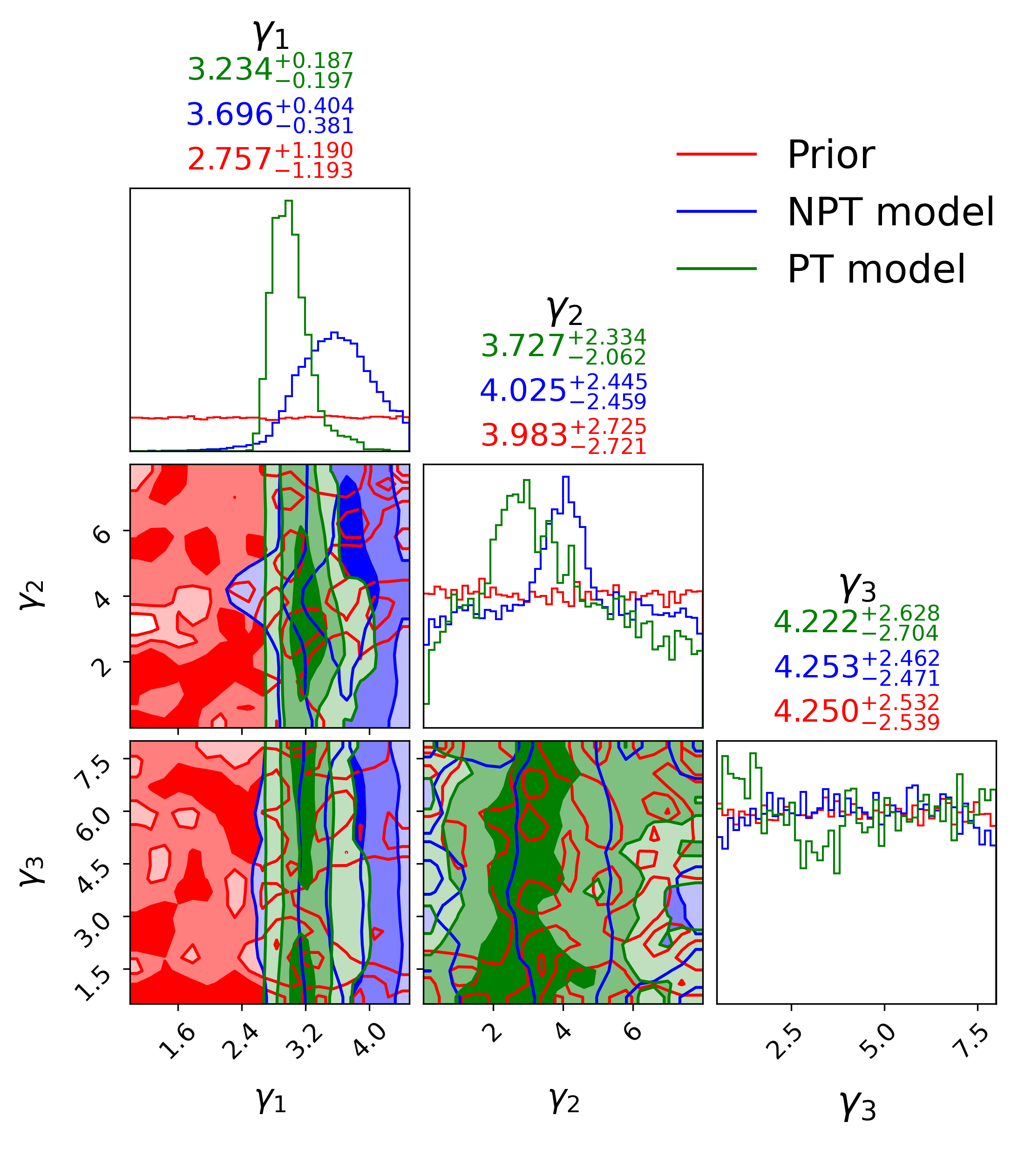}
    \caption{The posterior of the three polytrope indices after applying constraints to all existing observations with J0437$-$4715 included. Red is the prior, Blue is the indices posterior from the model without phase transition (NPT model), and green is the posterior from the model with phase transition included (PT model). The contour levels in the corner plot, going from deep to light colours, correspond to the 68\%, 84\%, and 98.9\% levels. The dashed line in the 1D corner plots represents the 68\% credible interval, and the title of this plot indicates the median of the distribution as well as the range of the 68\% credible interval.}
    \label{prior_post_compare_PT}
\end{figure}

\begin{figure}[htbp]
    \centering
    \includegraphics[scale=0.4]{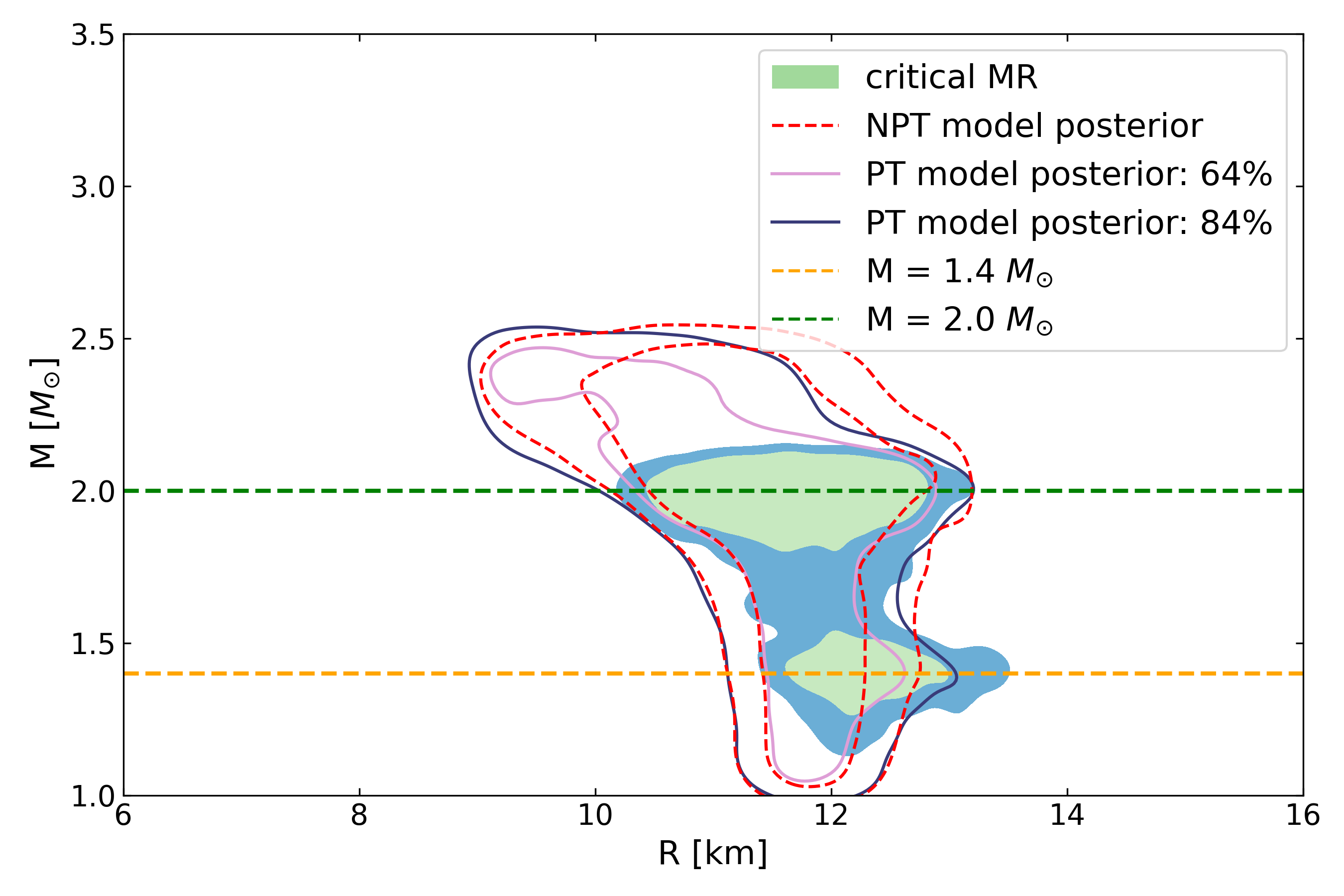}
    \caption{The MR posterior of the two models with and without phase transition. The red dashed region is for the model without phase transition, and the solid region is the model with phase transition. The solid contour represents the phase transition critical mass distribution of the PT model. The contour levels in the corner plot: the violet level is 64\%, and the dark blue level corresponds to 84\%.}
    \label{pt_MR}
\end{figure}

\begin{figure}[htbp]
    \centering
    \includegraphics[scale=0.4]{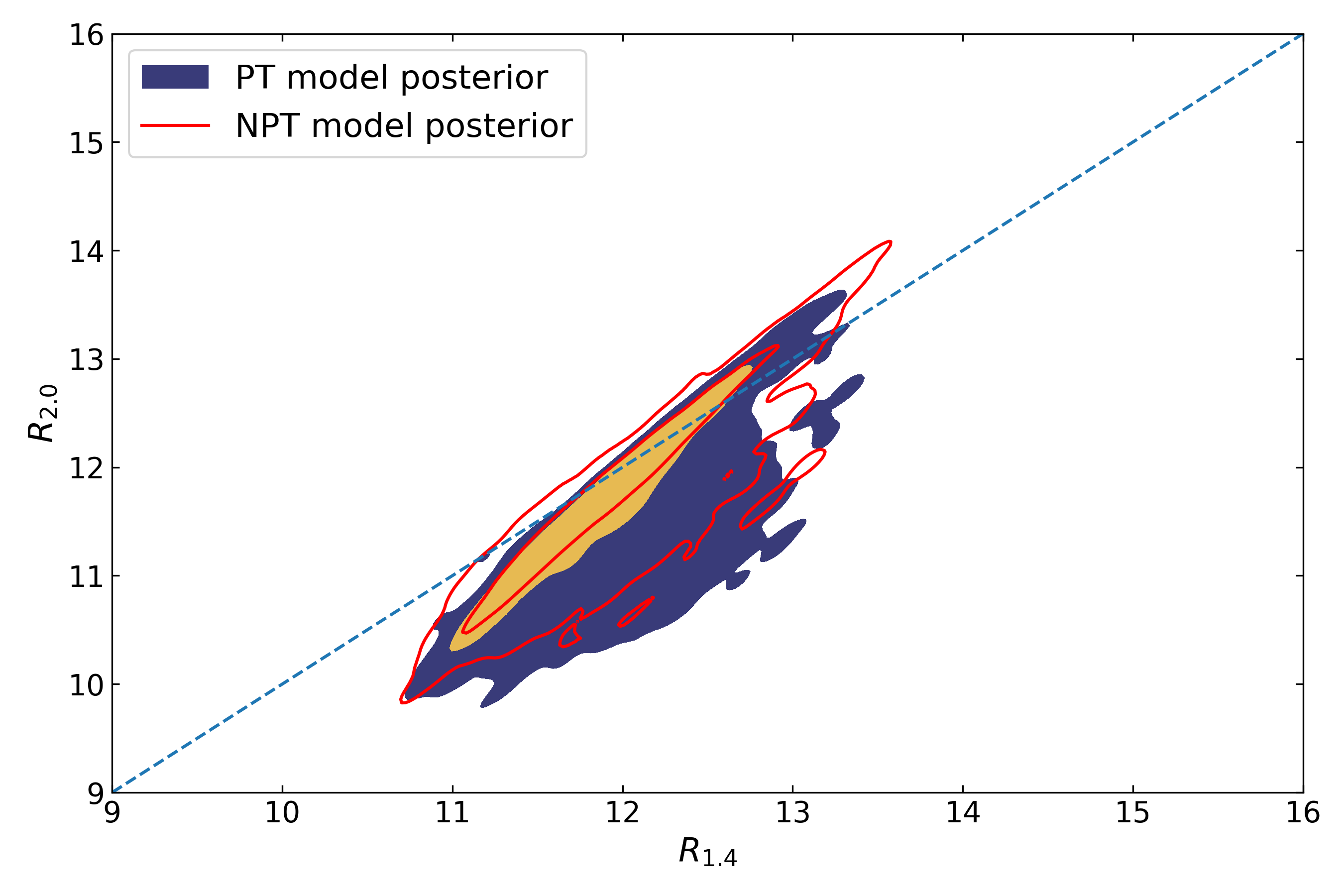}
    \caption{The contour of $R_{2.0}$ versus $R_{1.4}$. Posterior of the two models with and without phase transition. The red region is for the model without phase transition (NPT model), and the solid contour region is the model with phase transition (PT model). The contour levels in the corner plot, going from deep to light colours, correspond to the 68\% and 84\% levels.}
    \label{R1420}
\end{figure}

\section{Results of Phase Transition Constraint}\label{result}
In this section, we present the findings from our Bayesian inference analysis of phase transitions in neutron stars, with a particular focus on constraints derived from NICER observations. We begin by assessing the influence of the PSR J0437$-$4715 data on the inferred equation of state parameters, comparing the posterior distributions of key parameters in models with and without phase transitions. This allows us to evaluate the impact of including PSR J0437$-$4715 on the EOS constraints. Additionally, we examine the mass-radius (M-R) posterior distributions for both model types, discussing their implications for neutron star structure and the potential for phase transitions. Lastly, we utilized the Bayes factors to assess whether the observational data favor one model over the other.
\subsection{Constraint on Equation of State parameters}

After incorporating all current NICER observations, we plotted the posterior distributions of phase transition parameters with and without PSR J0437$-$4715 in Figure \ref{prior_post_compare}, along with the prior distributions. This figure illustrates the constraining power derived solely from NICER sources. Even without PSR J0437$-$4715, the previous NICER data already provide tight constraints on these phase transition parameters. However, including PSR J0437$-$4715 further tightens the constraints on the transition point $\varepsilon_t$, phase transition depth, $\Delta \varepsilon$ and the polytrope indices, especially $\gamma_1$. Specifically, the transition density shifts to the higher end, and the 68\% credible interval narrows compared to the constraints without PSR J0437$-$4715. The transition depth also shrinks to an even smaller range, for the 84\%  credible interval has changed. This reflects that the observation prefers a softer EOS and the transition point increase to maintain a suitable critical phase transition mass for neutron star so that this model can fit these two 1.4 $\msol$ stars (PSR J0437$-$4715 and PSR J0030+0451) separately.

This result is reasonable considering that PSR J0437$-$4715 has a similar mass but a different radius than PSR J0030+0451. Nevertheless, the overall improvement still needs to be improved due to the uncertainties associated with NICER measurements.

For the polytrope indices, we keep consistency with \cite{2013ApJ...773...11H}, a polytrope model with an added phase transition. We can safely compare this result with the NPT version of the model. Figure \ref{prior_post_compare_PT} shows the constraints on the polytrope indices from models with and without phase transitions. For $\gamma_1$, the model with phase transition exhibits a narrower posterior range and favors a smaller value, which is reasonable, as $\gamma_1$ determines the first segment of the polytrope before the phase transition. For accommodating the phase transition while satisfying the three NICER observations, a smaller $\gamma_1$ is required to soften the EOS, ensuring that the phase transition pressure does not become large. 

No significant constraints are put on these parameters for $\gamma_2$ and $\gamma_3$. In contrast, the $\gamma_2$ is shifted to a smaller value peak, reflecting that after the phase transition, the observations still prefer a softer EOS. However, no value has been excluded by NICER measurements, indicating that current observational precision still needs to be improved to constrain the high-density behavior of the EOS, which is consistent with previous findings.

The only difference between these two models is the inclusion of a phase transition. Thus, comparing the Bayesian evidence of these models is meaningful, as it provides crucial information on whether the observations favor the existence of a phase transition. The Bayes factor is the standard method in Bayesian statistics for comparing two models. In this analysis, the Bayes factor for comparing the NPT model with the PT model is defined as $B(\mathrm{NPT}/\mathrm{PT}) = \exp\left(\ln Z_1 - \ln Z_2\right) = Z_1/Z_2$
where $ Z_1 = p(\boldsymbol{d} \mid \mathrm{NPT}) $ represents the Bayesian evidence for the NPT model, and $ Z_2 = p(\boldsymbol{d} \mid \mathrm{PT})$ represents the Bayesian evidence for the PT model. In this context,$ \boldsymbol{d} $ denotes the observed dataset. Table \ref{Bayes_factor} lists the logarithms of the Bayesian evidence $\ln (Z) $for these two models, denoted as PT and NPT models, under two different observation data sets, without PSR J0437$-$4715 and this source. 

In our analysis, the Bayes' factor for the scenario that includes PSR J0437$-$4715 is 2.34, slightly smaller than 3.2, A Bayes' factor lower than 3.2 indicates that there is no clear evidence favoring one model over another. This suggests that current observations, even with the inclusion of new sources, still need to be more precise to distinguish between models with and without phase transitions definitively. However, we are very close to this threshold, and comparing this value with the Bayes' factor 1.40 without including PSR J0437$-$471, the improvement from this new observation PSR J0437$-$4715 appears to be crucial since it could increase the Bayes' factor significantly. By using Bayes Factor we see that our model shows that there is no clear preference for a polytrope model with or without a phase transition. However, the incorporation of new observational data can increase the Bayes factor in favor of the NPT model. this indicates that PSR J0437$-$4715 has the potential to identify phase transitions inside neutron stars, especially with its comparably high-precision radius measurement.

\subsection{Mass-Radius Posterior}

In \cite{Rutherford24}, the authors discuss the projection of the EOS posterior to the mass-radius (M-R) space, focusing on the predicted radius of 1.4 $M_{\odot}$, 2.0 $M_{\odot}$ star and the TOV mass of the polytrope EOS they are using. We extend that discussion by incorporating phase transitions into the polytrope EOS.

In Figure \ref{pt_MR}, we compare the M-R posterior distributions of models with and without phase transitions using two contours. The PT model exhibits a maximum mass comparable to the NPT model. Notably, the main feature of these two M-R posteriors is the appearance of pop-up regions around the 1.4 $\msol$ and 2.0 $\msol$ regions in the PT model, which tend to have larger radii in these mass ranges. Comparing the mass–radius posteriors presented in this work with those reported in \cite{Brandes:2023bob,PhysRevD.107.014011}, which also account for the phase transition, it is evident that the posteriors in the latter studies appear considerably smoother and exhibit fewer fluctuations than those in our plot. This discrepancy can be attributed to two primary factors. First, in \cite{Brandes:2023bob,PhysRevD.107.014011}, the mass–radius posteriors are derived directly from the MR curves without resampling the central density space for each EOS parameter set, which may result in a different representation of the mass–radius posterior. Second, in our analysis, the mass–radius posterior is computed based on EOS samples, and the observed fluctuations likely represent a minor artifact due to the limited sample size of the posterior.

To understand this, we computed the critical M-R for the PT model. The last stable star defines the critical M-R before the central density reaches the phase transition density, and the phase transition occurs within the neutron star. From the filled contour showing the critical M-R of these models, we can identify that the regions around 1.4 $\msol$ and 2.0 $\msol$ coincide with the central regions where phase transitions occur. These regions show a high correlation with the observations of interest.

For the 1.4 $\msol$ region, introducing the new observation PSR J0437$-$4715 is crucial. This star has a radius that slightly differs from PSR J0030$+$0451, necessitating a phase transition to reconcile these two observations. The first-order phase transition causes a slight mass increase, resulting in a more pronounced decrease in radius, which explains why the phase transition occurs in this region.

The other significant region is around the 2.0 $\msol$ star. The PT model must simultaneously fit the 1.4 $\msol$ and 2.0 $\msol$ observations. However, according to our observations, these two mass stars have very similar radii. Therefore, before reaching the 2.0 $\msol$ region, the stiff first piece of the polytrope EOS must become softer at the second piece to match the 2.0 $\msol$ radius measurement. This explains the emergence of this region.

The observed pop-up in the PT model reflects the fact that the occurrence of the phase transition softens the EOS. Consequently, the Bayesian technique tends to predict a larger radius for the critical star M-R than the NPT model, ensuring the model can still successfully explain the observations once softening occurs.

\begin{table}
\centering

\begin{tabular}{ccc}
\hline \hline 
\text{Fitted Model}  &\text{NPT model}& \text{ PT model} \\ 
\hline 
$R_{1.4}$     & 11.92$^{+1.08}_{-0.63}$      & 12.32$^{+0.95}_{-0.99}$\\
$R_{2.0}$ &  11.77$^{+1.21}_{-1.02}$    & 11.58$^{+1.37}_{-1.27}$\\
$M_{TOV}$ &  2.21$^{+0.25}_{-0.20}$      & 2.18$^{+0.28}_{-0.17}$  \\
$R_{TOV}$ &  11.19$^{+1.50}_{-1.65}$    & 11.01$^{+1.56}_{-1.73}$\\
\hline \hline
\end{tabular}
\caption{This table gives the 1.4 and 2.0 $\msol$ star radius (in km) from NPT and PT model, the unit for radius is km, for mass is $\msol$. Also we show the Maximum mass ($M_{TOV}$) predicted by NPT and PT model also the corresponding radius of maximum mass star($R_{TOV}$),The upper and lower values correspond to the 95\% credible interval}
\label{mtov_tab}
\end{table}

\section{Discussion of Implications}\label{discussion}
\begin{table*}[htbp]
  \centering
  \label{Table:4}
  
  \begin{tabular}{@{}lccc@{}}
    \hline\hline
    Threshold Values & Samples with PT & Percentage (\%) & Polytropic Index \\ 
    \midrule
    0.0001 & 0   & 0.00\%  & $\gamma_2$ \\
    0.001  & 24  & 0.04\%  & $\gamma_2$ \\
    0.01   & 73  & 0.12\%  & $\gamma_2$ \\
    \hline\hline
  \end{tabular}
  \caption{Number of samples that showed phase transition (out of 60,315 samples).}
  
\end{table*}
In \cite{Rutherford24}, the potential applications of the NPT model results for inferring the dense matter EOS and neutron star maximum masses are discussed, with a particular focus on predicting the radii of 1.4 $\msol$ and 2.0 $\msol$ stars, notice when the second piece of the polytrope in \cite{Rutherford24} take zero value, this model will back to a first-order phase transition included model. To deepen the discussion from the M-R posterior section, Figure \ref{critical} presents the radii of 1.4 and 2.0 $\msol$ stars predicted by both PT and NPT models that correspond to the intersection of the two dashed lines in Figure \ref{pt_MR}. Below, we will first discuss the application on 1.4 $M_{\odot}$ star radius and the radius difference between 2.0 $M_{\odot}$ star.

\subsection{Discussion on the 1.4 $\msol$ Star}

For the 1.4 $\msol$ star, as discussed in the M-R posterior section, the PT model exhibits a broader radius distribution. Comparing the 68\% credible intervals of the NPT and PT models reveals that they are very close, consistent with our observations in Figure \ref{pt_MR}. Within the same model, comparing the radii of 1.4 $\msol$ and 2.0 $\msol$ stars shows that the NPT model predicts them to be quite similar. This consistency is due to the need for a very soft EOS to fit both PSR J0437$-$4715 and PSR J0030+0451 while still explaining the observation of PSR J0740+6620, the 2.0 $\msol$ star. This requirement significantly shrinks the allowed equation of state parameter space since there is no first-order discontinuity along the Mass radius relation, while this is allowed for the PT model.

For the NPT model, the 2.0 $\msol$ star is found to have a smaller radius than the 1.4 $\msol$ star, which is due to the absence of a first-order phase transition and is a sign of stiff EOS. In Figure \ref{R1420}, we plot the contour of $R_{1.4}$ versus $R_{2.0}$ for both PT and NPT models and the line with a slope equal to 1. For samples with very large radii, $R_{2.0}$ is larger than $R_{1.4}$; otherwise, $R_{2.0}$ is always smaller. Comparing the PT and NPT models, more samples in the PT model have larger $R_{2.0}$, generally placing the PT model contour slightly above that of the NPT model in this diagram.

When comparing the difference between these two quantities, $\Delta R = R_{2.0} - R_{1.4}$, as shown in Figure \ref{Rdiff}, we observe that compared to the NPT model, the PT model has more samples in the smaller negative region. The mean value and 68\% credible interval are shifted to the lower side, which, according to \cite{PhysRevC.103.045808}, indicates the softening effect of the EOS within the neutron star. Here, this effect arises from the first-order phase transition. The NPT model has a more focused peak but still trends towards the negative side, consistent with the results of \cite{Rutherford24}. 

In \cite{Huang_2025}, an EOS-independent constraint on the radius of a \(1.4\,M_{\odot}\) neutron star was also discussed. For the ST+PDT model of J0030 adopted in this study, the EOS-independent analysis yielded \(R_{1.4} = 11.67^{+1.42}_{-1.50}\,\mathrm{km}\), which is broadly consistent with the NPT model. This agreement suggests that the relatively small \(R_{1.4}\) value may explain why the Bayes factor moderately favors the NPT model over the PT model. However, it is important to note that the EOS-independent modeling carries larger uncertainties because it relies solely on two NICER sources—PSR~J0030+0451 and PSR~J0437$-$4715—and does not incorporate additional multimessenger constraints. Consequently, these results are not yet conclusive for distinguishing between the NPT and PT models.

\subsection{Discussion on Neutron Star Maximum Mass}

Regarding the maximum mass of neutron stars, Figure \ref{mtov} illustrates that the TOV mass ($M_{TOV}$) range does not significantly differ between the two models. The PT model, however, displays a pronounced peak around the 2.0 $\msol$ region and generally predicts a smaller maximum mass. Additionally, the uncertainty band in $M_{TOV}$ is more significant for the PT model compared to the NPT model; however, given that the PT model has a sharper peak around 2.0 $\msol$, this TOV mass is instead more well-determined. This finding is logical because, although the maximum mass range of the PT model spans a broader range, it generally represents a softer EOS. Consequently, this model has more samples concentrated in the smaller maximum mass region due to the phase transition.

For specific values and credible intervals of the maximum mass, the radius of the 1.4 $\msol$ star, and 2.0 $\msol$ star, please refer to Table \ref{mtov_tab}. These quantities can be compared to the NICER group results as discussed in \cite{Rutherford24}. Our results are consistent with their findings within our credible intervals, especially for the same NPT model. Our results are broader than those in \cite{Rutherford24}, where the authors incorporated constraints from multi-messenger observations. In contrast, our analysis is limited to constraints derived solely from X-ray observations.

\subsection{Discussion on the Without Phase Transition (NPT) model}

In the NPT model, we observed that for certain samples of mass and radius, there were instances where we observed phase transitions. The NPT model uses a piecewise polytropic equation of state segmented into three distinct regions, each having its own polytropic index ($\gamma_1$, $\gamma_2$, $\gamma_3$) and transition densities ($\epsilon_1$, $\epsilon_2$). A key signature of a first-order phase transition implemented via a Maxwell construction is that the speed of sound drops to zero within the coexistence interval. In our analysis, this behavior is effectively captured when any polytropic index approaches zero. Due to the constraints imposed by our prior ranges, only $\gamma_{2}$ has the potential to approach arbitrarily close to zero. When $\gamma_{2} = 0$, it signifies that the speed of sound also drops to zero, serving as the precise indicator we use to identify phase transitions. The occurence of phase transitions within the NPT model, is a consequence of the model parameterization method and the prior ranges chosen. Specifically, the detection of phase transitions in certain samples at different threshold values arises because one of the polytropic indices ($\gamma_2$) approaches zero, resulting in a highly compressible segment that behaves as a phase transition. The chosen prior ranges, allow one of the three polytropic indices to have values close to zero while the others remain within physically plausible bounds. We had tested our NPT model with different threshold values. In total, we had 60,315 samples. For a tight threshold value of 0.0001 none (0\%) of the samples showed phase transition. However, when we softened the threshold value to 0.001 and 0.01 we found that 0.04\% and 0.12\% of the total samples showed phase-transitions like behavior in this NPT model. We have also shown this in Table \ref{Table:4}.
{The relatively infrequent occurrence of this $\gamma_2$ approaching to 0 behavior serves as a benchmark for comparison with the Bayesian inference results, suggesting that a phase transition may not be critically required at least with current observation.}
\section{Conclusions}\label{conclusion}

In conclusion, this research enhances the understanding of neutron star interiors by exploring the constraints on phase transition parameters using NICER observations, focusing on the new source PSR J0437$-$4715. The results indicate that incorporating this new result tightens the constraints on the phase transition density and depth, shifting the transition density to a higher range. The Bayes factor between PT and NPT models reveals a slight increase of the Bayes' factor for preferring models equipped without phase transitions. However, the precision of current observations is still insufficient for any substantial conclusion. The M-R posterior analysis highlights the significant impact of phase transitions on predicted radii, particularly around 1.4 and 2.0 $\msol$. This finding aligns with the necessity of a softer equation of state to simultaneously fit multiple observational constraints. We also identified phase-transition-like behavior in the standard polytropic model, even without an explicitly implemented phase transition. Since the phase-transition-like behavior is rare in this NPT model, this result provides a reference point for the Bayesian inference results, implying that a phase transition may not be critically necessary based on current observations. The study also suggests that while the maximum mass predictions are similar between models with and without phase transitions, the uncertainties are more significant for phase transition models, reflecting the inherent complexities of these models. Overall, the inclusion of PSR J0437$-$4715 in the analysis presents a promising avenue for further refining the constraints on neutron star interiors, emphasizing the need for more precise and extensive observational data to conclusively determine the presence and characteristics of phase transitions within neutron stars.

\section*{Acknowledgements}

C.H., S.S acknowledges support from an Arts \& Sciences Fellowship at Washington University in St. Louis, C.H. also aknowleges support from NASA grant 80NSSC24K1095. The author extends gratitude to  Anna Watts, Devarshi Choudhury and Nathan Rutherford for their valuable suggestions and comments, which significantly improved this manuscript.

\bibliography{sample631}{}
\bibliographystyle{aasjournal}

\end{document}